\documentclass{article}
\usepackage{graphicx,psfig,epsfig}
\def\title{\begin{center}\Large\bf}
\def\author(s){\vspace{0.3cm}\large\rm}
\def\text{\end{center}}

\newcommand{\oiii}{[O~{\sc iii}]~5007~\AA}
\newcommand{\arcmin}{$^{\prime}$~}
\newcommand{\arcsec}{$^{\prime\prime}$~}
\begin{document}

\title
Optical Survey of Planetary Nebulae in the northern Galactic Bulge

\author(s)
P. Boumis$^{\rm 1,2}$ and J. Papamastorakis$^{\rm 1,2}$

$^{\rm 1}${\it University of Crete, P.O Box 2208, GR-71003, Heraklion, Greece}\\
$^{\rm 2}${\it Foundation for Research and Technology-Hellas, P.O. Box 1527,\\
GR-71110 Heraklion, Greece}\\
\text

\vspace{0.3cm}

\large

\section*{Abstract}
We present preliminary results of an optical survey for Planetary
Nebulae (PNe) in the northern part of the Galactic Bulge. In
particular, observations at wavelengths corresponding to the [O{\sc iii}]
5007 emission line have been carried out with the 0.3 m telescope at
Skinakas Observatory, where we have detected several new PNe and have
rediscovered known and known-possible PNe from that region.

\section{Introduction}
The Galactic Planetary Nebulae (PNe) are always under high
investigation because of our difficulties to determine their
properties (like their number, distance, chemical composition,
morphology and kinematics), which play a very important role to the
chemical enrichment history of the interstellar medium as well as to
the star formation and evolution of our Galaxy. The Galactic Bulge is
covered by the region of Galactic longitude $l = \pm 15^{o}$ and
Galactic latitude $b = \pm 10^{o}$ (dimensions $\sim$2 Kpc toward and
$\sim$1.5 perpendicular the Galactic plane - Weiland et al. 1993). In
total, the Bulge contains about 10$^{11}$ stars of Population II as
well as gas ($\sim 10^{6} {\rm M}_{\odot}$) and dust (mass $\sim$100
times less than the mass of the gas - Morison \& Harding 1993). It is
generally accepted that a big number of PNe exist in the Galactic
Bulge, but because of the interstellar extinction, only a small number
of them (less than 300) have been discovered (K\"{o}ppen \& Vergely
1998). According to Acker et al (1992a) and Zijlstra \& Pottasch
(1991), the total number of PNe in the Galaxy can be estimated between
15,000 to 30,000. Up to date, less than 2,500 PNe have been discovered
(Acker et al. 1992b, 1996), which means that only a small amount of the
total PNe have already been detected. In a very recent catalogue,
Kohoutek (2001) present an additional 169 new PNe which were
discovered through different surveys between the years 1995-1999.

Systematic \oiii~surveys have already been done in the past to
identify PNe in other galaxies, and their results were very promising,
since $\sim$1600 new PN detection were made (Feldmeier et al. 1997 and
references therein). The above results show that the right combination
of the telescope size and the CCD camera, which have been used for PN
detections at different distances, result to a lot of new
discoveries. Therefore, we suggest that a large number of new PNe can
be detected in the Galactic Bulge by using a 0.3 m telescope of large
field of view (combined with a high quality CCD camera) and a narrow
\oiii~filter. We must note that a PN discovery has already been done
in the past using this configuration (Xilouris et al. 1994). Take in
account the fact that a PNe survey in the bulge has never been done in
the past using this emission line except from a limited number of
images, the \oiii~emission line was chosen for our survey.

\section{Observations}
The observations were made with the 0.3 m Schmidt-Cassegrain (f/3.2)
telescope at Skinakas Observatory in Crete, Greece. In our survey we
decided to observe the regions $10^{o} < l < 20^{o}$, $-10^{o} < b <
-3^{o}$ and $0^{o} < l < 20^{o}$, $3^{o} < b < 10^{o}$ (see
Fig. 1). The reasons to do that was (a) the site of the Observatory
which allow us to observe down to $-25^{o}$ declination and (b) the
high interstellar extinction in the \oiii~emission line between
$-3^{o} < b < 3^{o}$. In Fig. 1, the optical imaging survey grid in
equatorial coordinates is shown. Galactic coordinates are also
included (dash lines), while the small empty rectangles represent the observed
fields. We must note that we covered at first a spesific region of the
proposed grid to make sure that our discovery method works well. 
Two exposures in \oiii~of 1200 s and three exposures in the nearby Continuum Red of 180 s were taken to make sure that the cosmic rays will be removed
correctly.

\begin{figure}
\centering
\mbox{\epsfclipon\epsfxsize=3.5in\epsfbox[0 0 488 434]{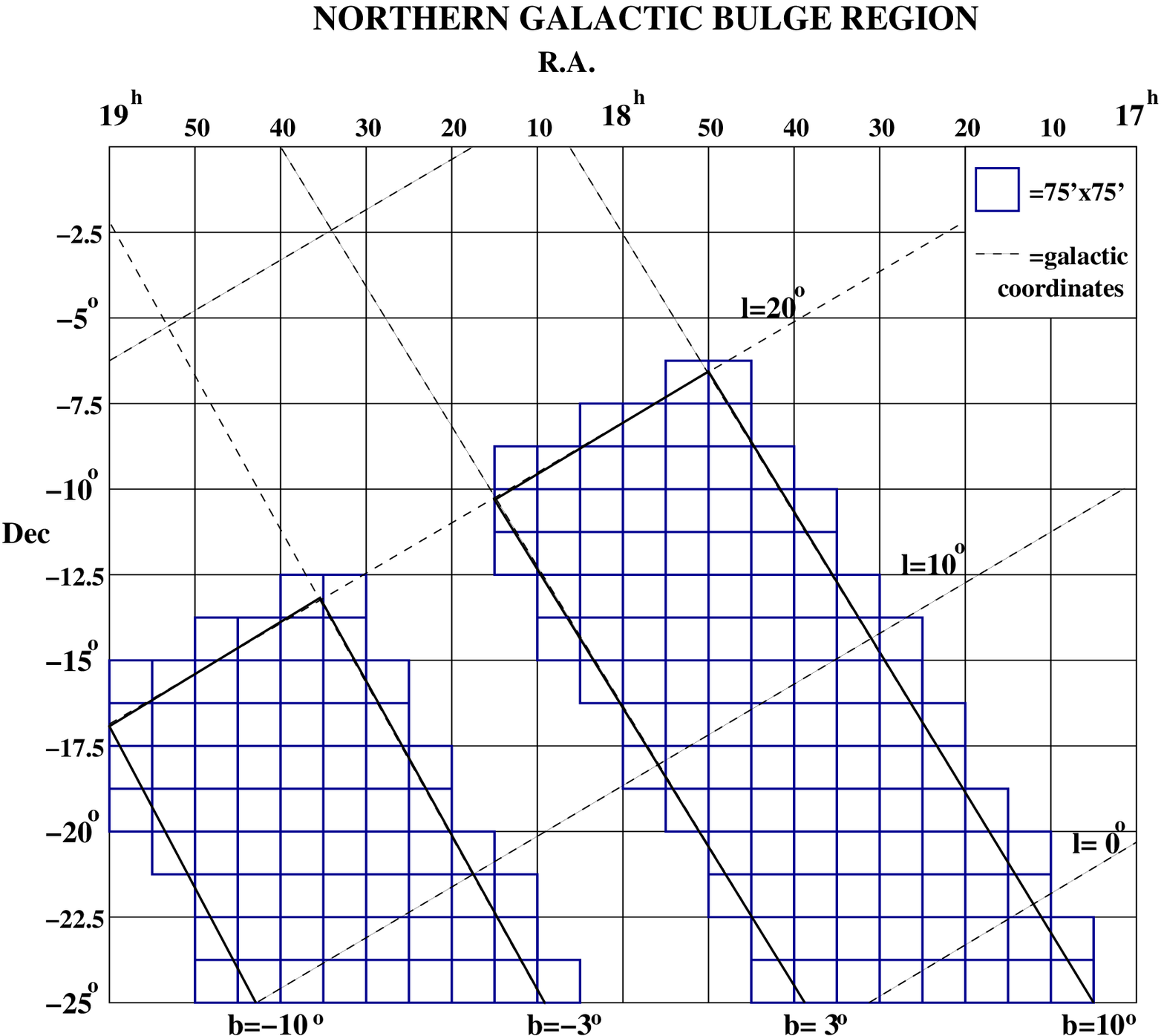}}
\\ {{\bf Figure 1.} Optical imaging survey grid in equatorial
coordinates. Galactic coordinates are also including (dash lines) to
permit an accurate drawing of the selected Bulge region (bold solid
lines). The small empty rectangles represent the observed fields.}
\label{fig1}
\end{figure}

\section{Detection method}
Our detection method is similar but not the same to that followed by
Beaulieu et al. (1999). In particular, after having the correct
scaling, we subtracted both the two [O{\sc iii}] images by the
corresponding continuum image for each field. Then, the initial PN
identification was performed by ``blinking'' each subtracted [O{\sc
iii}] frame by the continuum. Given that our pixels have an angular
size of $\sim$ 4 arcsec if projected on the sky, the PN candidates at
the bulge appeared as point (or sometimes extended) positive sources
on the [O{\sc iii}] images, but were absent on the continuum. As it
was mentioned above, the resulting positive sources could be a PN
candidate as well as a cosmic ray. To make sure that no cosmic rays
were included among the candidates, we checked individually the [O{\sc
iii}] frames for each field. Objects that appeared bright on the first
frame but they were invisible on the second, counted as cosmic rays
because a real emission source must be appeared in both the [O{\sc
iii}] images. The procedure of how the PN candidates were look like
through the different steps can be seen in Fig. 2. The first set of
images (a) is a known PN which was found and presented for comparison
reasons; the other set (b) is new discovery.
 
After finding the PN candidates, the next step was to perform an
astrometric solution for all the images containing one or more
candidates. In order to calculate the equatorial (and Galactic)
coordinates of the PN candidates we used the Hubble Space Telescope
(HST) Guide Star Catalog (Lasker et al. 1999) and IRAF routines (Image
package). The coordinates for the PN candidates were then
calculated. Based on the stars of our frames and the HST Guide stars,
we also estimate the error in one frame astrometry to be less than
1.2\arcsec. The new objects were then checked in order to identify the
already known PNe. For that purpose we used any up-to-date published
catalogue which is related to planetary nebulae. In the near future
(Boumis et al. 2001) detailed results (imaging and spectroscopy) for
the newly discovered PNe found in our survey will be presented.

\section*{Acknowledgments}
PB acknowledges support from a "P.EN.E.D." 
program of the General Secretariat of Research and Technology of
Greece. We also thank E. Semkov for his help during the
observations. Skinakas Observatory is a collaborative project of the
University of Crete, the Foundation for Research and Technology-Hellas
and the Max-Planck-Institut fur Extraterrestrische Physik.

\section*{References}
Acker A., Cuisinier F., Stenholm B. \& Terzan A., 1992a, A\&A, 264, 217\\
Acker A., Ochsenbein F., Stenholm B., Tylenda R., Marcout J. \& Schohn
C., 1992b, in the Strasbourg-ESO Catalogue of Galactic Planetary
Nebulae, Parts 1 and 2 (Strasbourg: ESO)\\
Acker A., Marcout J., Ochsenbein F., Beaulieu S., Garcia-Lario P. \&
Jacoby G., 1996, First Supplement to the Strasbourg-ESO Catalogue of
Galactic Planetary Nebulae (Strasbourg: Obs. Strasbourg)\\
Beaulieu S. F., Dopita M. A. \& Freeman K. C., 1999, ApJ, 515, 610\\
Boumis P., Paleologou E. V., Mavromatakis F. \& Papamastorakis J.,
2001, MNRAS, to be submitted\\
Feldmeier J. J., Ciardullo R. \& Jacoby G. H., 1997, ApJ, 479, 231\\
Kohoutek L., 2001,in the Catalogue of Galactic Planetary Nebulae -
Version 2000, Abhandl. Hamburger Sternwarte XII. (CGPN (2000))\\
K\"{o}ppen J. \& Vergely J.-L., 1998, MNRAS, 299, 567\\
Lasker B. M., Russel J. N. \& Jenkner H., 1999, in the HST Guide Star
Catalog, version 1.1-ACT, The Association of Universities for Research
in Astronomy, Inc\\
Morison H. L. \& Harding P., 1993, PASP, 105, 977\\
Weiland J. L., Arendt R. G., Berriman G. B., et al., 1994, ApJ, 425, L81\\
Xilouris K. M., Papamastorakis J., Sokolov N., Paleologou E. \& Reich
W., 1994, A\&A, 290, 639\\
Zijlstra A. A. \& Pottasch S. R., 1991, A\&A, 243, 478

\begin{figure} 
\centering
\mbox{\epsfclipon\epsfxsize=4.5in\epsfbox[19 19 429 797]{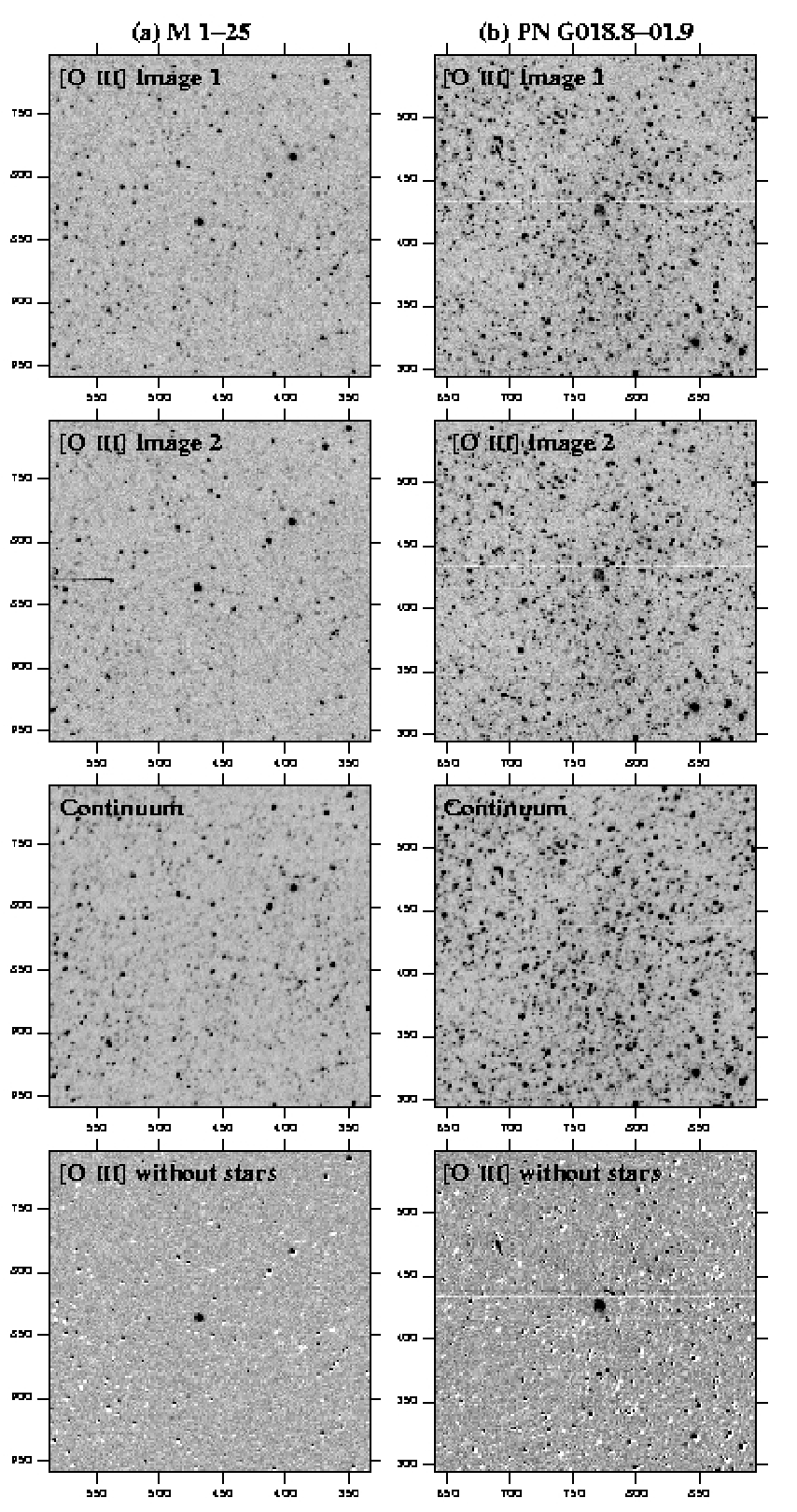}}
\\ {{\bf Figure 2.} Examples of images (17\arcmin on both sides each)
for two different objects which show our survey's typical
discoveries. The first set of images (a) is a known PN which was found
and presented for comparison reasons; the other set (b) is a new
discovery. Note that the remaining black dots in the last image of
each set (except the PNe) are cosmic rays and stars which did not
remove with the continuum subtruction. North is at the top, East to
the left.}
\label{fig2}
\end{figure}

\end{document}